\documentclass[aps,prl,superscriptaddress,twocolumn,longbibiliography]{revtex4-2}

\usepackage{ulem} 

\usepackage[hidelinks]{hyperref}
\hypersetup{
	colorlinks,
	linkcolor={red},
	citecolor={blue},
	urlcolor={blue}
}
\usepackage{physics}
\usepackage{balance}
\usepackage{amssymb}
\usepackage{suffix}
\usepackage{mathtools}
\usepackage[utf8]{inputenc}
\usepackage{booktabs}
\usepackage{cases}
\usepackage[multiple]{footmisc}
\usepackage{dcolumn}
\usepackage{color,soul}
\usepackage{rotating}
\usepackage{perpage}
\usepackage{siunitx}
\usepackage{xcolor}
\usepackage{soul}
\usepackage{amsmath}
\usepackage{tikz}
\usepackage[T1]{fontenc}
\usepackage{etoolbox}
\usepackage{graphics}
\usepackage{siunitx}
\usepackage{float}	
\usepackage{collref}
\usepackage{multirow}
\usepackage{mathtools}
\usepackage{bm}
\usepackage{url}

\usepackage{tikz}
\usepackage{tikz-3dplot}
\usepackage{accents}

\makeatletter
\newcommand{\doublewidetilde}[1]{{%
		\mathpalette\double@widetilde{#1}%
}}
\newcommand{\double@widetilde}[2]{%
	\sbox\z@{$\m@th#1\widetilde{#2}$}%
	\ht\z@=.9\ht\z@
	\widetilde{\box\z@}%
}
\makeatother

\usepackage{etoolbox,lipsum}

\newcommand{\blue}[1]{\textcolor{blue}{#1}}

\begin{document} 
	\title{Spin polarization engineering in $d$-wave altermagnets}
	
	\author{Mohsen Yarmohammadi}
	\email{mohsen.yarmohammadi@georgetown.edu}
	\affiliation{Department of Physics, Georgetown University, Washington DC 20057, USA} 
	\author{Marco Berritta}
	\affiliation{Department of Physics and Astronomy, Uppsala University, Box 516, SE-75120 Uppsala, Sweden}
	\author{Marin Bukov}
	\address{Max Planck Institute for the Physics of Complex Systems, N\"othnitzer Str.\ 38, 01187 Dresden, Germany}
	\author{Libor \v{S}mejkal}
	\address{Max Planck Institute for the Physics of Complex Systems, N\"othnitzer Str.\ 38, 01187 Dresden, Germany}
	\address{Max Planck Institute for Chemical Physics of Solids, N\"othnitzer Str.\ 40, 01187 Dresden, Germany}
	\address{Institute of Physics, Czech Academy of Sciences, Cukrovarnick\'a 10, 162 00 Praha 6, Czech Republic}
	\author{Jacob Linder}
	\affiliation{Center for Quantum Spintronics, Department of Physics, Norwegian University of Science and Technology, NO-7491 Trondheim, Norway}
	\author{Peter M. Oppeneer}
	\affiliation{Department of Physics and Astronomy, Uppsala University, Box 516, SE-75120 Uppsala, Sweden}
	\date{\today}
	\begin{abstract}
		Altermagnets host unconventional spin-polarized bands despite zero net magnetization, but controlling their spin structure remains challenging. We propose a multi-field approach to engineer spin polarization in $d$-wave altermagnets using gating, optical driving, and in-plane electric fields, which enable tunable and switchable polarizations along multiple directions. Optical driving induces out-of-plane ($z$) polarization, while gating and in-plane fields generate $x$- and $y$-polarizations via the Edelstein effect, all of which are experimentally detectable. We further find that spin- and band-selective doping induces chiral optical activity, a feature unique to altermagnets. Our approach provides a versatile route for full control of spin polarization in altermagnets.		
	\end{abstract}
	\maketitle
	{\allowdisplaybreaks
		\blue{\textit{Introduction.}}---Altermagnets are a new class of magnetic materials~\cite{PhysRevX.12.040501,PhysRevX.12.031042,PhysRevX.12.040002,Krempaský2024} that break the traditional division between ferromagnets and antiferromagnets and exhibit unconventional properties~\cite{PhysRevX.12.040501,PhysRevX.12.031042,PhysRevX.12.040002,Krempaský2024,doi:10.1126/sciadv.aaz8809,PhysRevB.99.184432,Naka2019,doi:10.7566/JPSJ.88.123702,PhysRevX.12.011028,PhysRevLett.128.197202,PhysRevLett.130.216701,PhysRevLett.134.196703}. Despite zero net magnetization, their spin symmetries and nonrelativistic electronic structure yield symmetry-protected, alternating spin polarization in momentum space. Spin-momentum locking can appear in $d$-, $g$-, or $i$-wave forms~\cite{PhysRevX.12.040501,PhysRevX.12.040002,PhysRevX.12.031042,PhysRevB.99.184432,doi:10.7566/JPSJ.88.123702,Reimers2024}, with $d$-wave altermagnets predicted to host unusual current-driven spin effects, including exotic spin torques, spin-orbit torques, and spin-splitter currents~\cite{Bose2022,PhysRevLett.128.197202,PhysRevLett.129.137201,PhysRevLett.126.127701,doi:10.1126/sciadv.adn0479,Jiang2025}.
		
		Initial experimental indications of altermagnetic properties and functionalities were reported~\cite{PhysRevB.102.075112,PhysRevB.103.125114,Feng2022,Reichlova2024,PhysRevB.109.094413}, however, precise control of their spin polarization remains a key challenge for spintronics. Previous static approaches—strain, doping, or (anti)ferroelectric tuning~\cite{khodas2025tuningaltermagnetismstrain,PhysRevB.109.144421,Li2025,8zlt-mlms,león2025strainenhancedaltermagnetismca3ru2o7,wickramaratne2025effectsaltermagneticorderstrain,PhysRevLett.134.106801,PhysRevLett.134.106802,šmejkal2024altermagneticmultiferroicsaltermagnetoelectriceffect}—can face limits such as lattice instabilities and poor tunability, while dynamical methods provide greater flexibility~\cite{PhysRevB.79.081406,Rudner2020,RevModPhys.89.011004,Xu2025,yarmohammadi2025anisotropiclighttailoredrkkyinteraction,ghorashi2025dynamicalgenerationhigherorderspinorbit,rajpurohit2024opticalcontrolspinsplittingaltermagnet}. Specifically, Floquet engineering, using periodic driving to modify quantum systems~\cite{PhysRevB.79.081406,Bukov2015,RevModPhys.89.011004,Rudner2020}, can reshape band structures and spin textures~\cite{PhysRevResearch.4.033213,Sentef2015,PhysRevLett.121.080401,10.1093/ptep/ptad007}, enabling non-equilibrium spin transport.

		The spin Edelstein effect~(generation of a non-equilibrium spin polarizations from an applied electric field in systems with spin-orbit coupling~(SOC))~\cite{EDELSTEIN1990233,PhysRevLett.118.116801,Lesne2016,RevModPhys.91.035004} usually relies on SOC and spin-momentum locking, while the orbital Edelstein effect can occur even without SOC~\cite{Salemi2019,PhysRevB.110.L201407,Johansson_2024,PhysRevResearch.5.043294,edelstein1990spin,PhysRevMaterials.5.074407}. Recently, an out-of-plane electric field was shown to generate altermagnetic spin polarization in certain monolayer antiferromagnets~\cite{mazin2023inducedmonolayeraltermagnetismmnpsse3}. Additionally, various current-induced spin polarization effects were proposed, including relativistic variants in altermagnetic interfaces~\cite{trama2024nonlinearanomalousedelsteinresponse}, nonrelativistic variants in bulk altermagnets~\cite{golub2025spinorientationelectriccurrent}, $p$-wave antialtermagnets~\cite{hellenes2024pwavemagnets,Jungwirth2025,Chakraborty2025}, and chiral spin-textured magnets~\cite{Hu2025}.
        
		In this Letter, we propose a \textit{multi-field platform}, absent in the previous proposals, for more complete control of spin polarization in two-dimensional~(2D) $d$-wave altermagnets. The method unifies electrostatic gating, optical driving, and in-plane electric fields. Gating induces Rashba spin-orbit coupling (RSOC)~\cite{winkler2003spin,yarmohammadi2025anisotropiclighttailoredrkkyinteraction,PhysRevB.110.054427,ghorashi2025dynamicalgenerationhigherorderspinorbit,mazin2023inducedmonolayeraltermagnetismmnpsse3}, while circularly polarized light (CPL) generates anisotropic spin textures and spin polarization along $z$-direction~\cite{yarmohammadi2025anisotropiclighttailoredrkkyinteraction,ghorashi2025dynamicalgenerationhigherorderspinorbit}. In addition, in-plane electric fields produce spin Edelstein polarizations along $x$- and $y$-directions. Proposed materials include thin films of $d$-wave altermagnets KV$_2$Se$_2$O~\cite{Jiang2025}, RbV$_2$Te$_2$O~\cite{Zhang2025}, RuO\(_2\)~\cite{Feng2022,doi:10.1126/sciadv.aaz8809,PhysRevLett.128.197202,weber2024opticalexcitationspinpolarization}, and $\kappa$-CL~\cite{Naka2019}, noting RuO\(_2\) remains debated.
       
       Without CPL, gating and in-plane electric fields generate spin Edelstein polarizations of opposite sign along $x$- and $y$-directions. Adding CPL breaks this reversal symmetry, yielding strongly anisotropic polarizations. We demonstrate large, tunable, and switchable spin polarizations enabled by light-controlled band topology~\cite{Smejkal2018} and symmetry breaking, offering a route to full spin control in altermagnets. Intrinsic altermagnetic order can also generate chiral optical activity under spin-selective doping, opening routes to spin-resolved photonic technologies~\cite{deriy2025opticalspintronicsopticalcommunication,Zhang2025,x10.1063/1.4945715}.

		\blue{\textit{Effective Hamiltonian model.}}---We consider a 2D Rashba $d$-wave altermagnet, as illustrated in Fig.~\ref{f1}(a), for both $d_{x^2 - y^2}$ and $d_{xy}$ pairing symmetries~\cite{reichlova2021macroscopictimereversalsymmetry}, described by the Hamiltonian{\small\begin{align}
				\mathcal{H}^{\rm R}_{\vec{k}} =  \frac{\hbar^2 (k_x^2 + k_y^2)}{2m_{\rm e}} \sigma_0 
				+ h(\vec{k})\, \sigma_z + \lambda\left(k_x \sigma_y - k_y \sigma_x \right)\,,
		\end{align}}where $h(\vec{k}) = \frac{\hbar^2\mathcal{D}(k_x^2 - k_y^2)}{2m_{\rm e}} $ or $\frac{\hbar^2 \widetilde{\mathcal{D}}k_x k_y}{m_{\rm e}} $, with $m_{\rm e}$ the electron mass. The parameters $0 < \mathcal{D}\,{\rm or}\,\widetilde{\mathcal{D}} \leq 1$ characterize the strength of the $d_{x^2 - y^2}$ or $d_{xy}$ altermagnets, respectively, while $\vec{\sigma}~(\sigma_0)$ represents the Pauli vector~(identity matrix).
	
	Top and bottom gate electrodes with voltages $\pm V$ are included to emulate a realistic device setup. They introduce RSOC with strength $\lambda$, which couples spin and momentum, breaks inversion symmetry, and modifies spin splitting~\cite{winkler2003spin,yarmohammadi2025anisotropiclighttailoredrkkyinteraction,PhysRevB.110.054427}.\begin{figure}[t]
	 	\centering
	 	\includegraphics[width=0.95\linewidth]{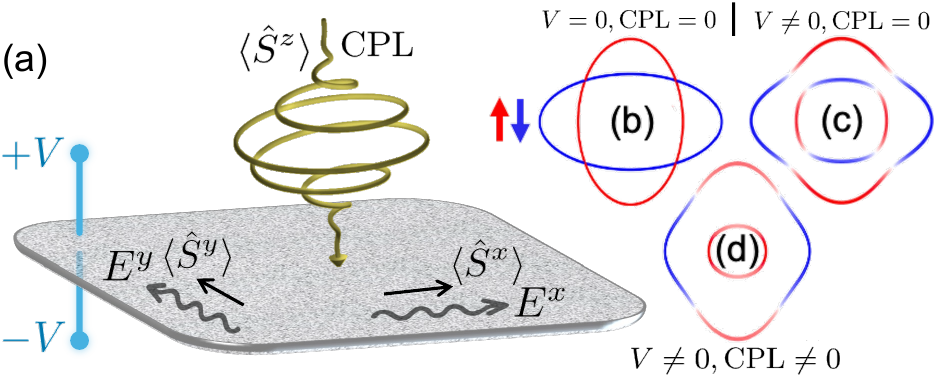}
	 	\caption{(a) Schematic of a two-dimensional $d$-wave altermagnet gated by $\pm V$ electrodes and illuminated by circularly polarized light (CPL, gold colored spiral). (b) Spin-resolved Fermi contours for pristine spin-up (red) and spin-down (blue) states. (c) Gating induces Rashba spin-orbit coupling, reshaping the Fermi surface while maintaining spin balance. (d) CPL breaks this balance, generating anisotropic spin textures. CPL engineers out-of-plane polarization $\langle \hat{S}^z \rangle$ via Floquet drives, while gating and in-plane fields $(E^x,E^y)$ allow tunable and switchable control of in-plane \(\langle \hat{S}^x \rangle\) and \(\langle \hat{S}^y \rangle\) via the Edelstein effect. The spin polarizations are noncollinear.}
	 	\label{f1}
	 \end{figure}  
		
		We next introduce a time-periodic vector potential for right-handed CPL, $\vec{A}(t) = A_{\rm d} [\sin(\Omega_{\rm d} t), \cos(\Omega_{\rm d} t)]$, with driving frequency $\Omega_{\rm d}$ and amplitude $A_{\rm d}$. Minimal coupling $\vec{k} \to \vec{k} - \frac{e}{\hbar} \vec{A}(t)$ makes the Hamiltonian time-periodic, $\mathcal{H}^{\rm R}_{\vec{k}(t)}(t+T) = \mathcal{H}^{\rm R}_{\vec{k}(t)}(t)$ with $T = 2\pi/\Omega_{\rm d}$ (see Appendix~\hyperlink{mylinkA}{A}). 
		
		In the high-frequency regime ($\hbar \Omega_{\rm d} \approx 1$ eV), an effective time-independent Floquet Hamiltonian is obtained via van Vleck expansion~\cite{yarmohammadi2025anisotropiclighttailoredrkkyinteraction,PhysRevB.84.235108,PhysRevLett.110.026603,PhysRevB.107.054439,PhysRevB.111.014440}:\begin{align}
			\mathcal{H}^{\rm eff}_{\vec{k}} \approx \mathcal{H}_0^{\rm F}(\vec{k}) + \frac{[\mathcal{H}^{\rm F}_{-1}(\vec{k}), \mathcal{H}^{\rm F}_{+1}(\vec{k})]}{\hbar\Omega_{\rm d}},
			\end{align}with Floquet components $\nu = \{-1,0,+1\}$:\begin{align}
				\mathcal{H}^{\rm F}_{\nu}(\vec{k}) = {} \frac{1}{T} \int_0^T \mathcal{H}^{\rm R}_{\vec{k} - \tfrac{e}{\hbar}\,\vec{A}(t)}(t) e^{ i \nu \Omega_{\rm d} t} dt\, .
			\end{align}We define \(\mathcal{A} = eA_{\rm d}/\hbar\) and light-induced exchange potential $\Delta = \lambda^2 \mathcal{A}^2/\hbar \Omega_{\rm d}$. To leading order, we obtain the effective Hamiltonians{\small\begin{widetext}\begin{subequations}
					\begin{align}
						\mathcal{H}^{{\rm eff},d_{x^2-y^2}}_{\vec{k}} \! &\approx \! \bigg( \frac{\hbar^2 (k_x^2 + k_y^2)}{2m_{\rm e}} +\frac{\hbar^3\,\Omega_{\rm d}\Delta}{2\,m_{\rm e}\,\lambda^2} \bigg) \sigma_0 
						+ \! \bigg( \frac{\hbar^2 \mathcal{D} (k_x^2 - k_y^2)}{2m_{\rm e}}+\Delta \bigg) \sigma_z+ \lambda\left[\left( 1 - \frac{\hbar^2\, \Delta\,\mathcal{D}}{m_{\rm e}\,\lambda^2} \right)k_x \sigma_y 
						- \! \left( 1 + \frac{\hbar^2\, \Delta\,\mathcal{D}}{m_{\rm e}\,\lambda^2} \right)k_y \sigma_x \right],\label{eq_2a}\\
						\mathcal{H}^{{\rm eff},d_{xy}}_{\vec{k}}\! &\approx {} \! \bigg( \frac{\hbar^2 (k_x^2 + k_y^2)}{2m_{\rm e}} +\frac{\hbar^3\,\Omega_{\rm d}\Delta}{2\,m_{\rm e}\,\lambda^2} \bigg) \sigma_0 + \! \bigg( \frac{\hbar^2 \widetilde{\mathcal{D}} \,k_x\, k_y}{m_{\rm e}}+\frac{\Delta}{2} \bigg) \sigma_z + \! \lambda\left[\left( 1 - \! \frac{\hbar^2\, \Delta\,\widetilde{\mathcal{D}}}{2\,m_{\rm e}\,\lambda^2}\tfrac{k_y}{k_x} \right) k_x\sigma_y 
						- \! \left( 1 - \frac{\hbar^2\, \Delta\,\widetilde{\mathcal{D}}}{2\,m_{\rm e}\,\lambda^2}\tfrac{k_x}{k_y} \right) k_y \sigma_x \right].
		\end{align}\end{subequations}\end{widetext}}In both Hamiltonians,
        the first term proportional to \(\Omega_{\rm d} \Delta \sigma_0\) leads to a uniform energy shift---chemical potential. An additional constant spin splitting arises from the term \(\Delta \sigma_z \)~(i.e., light-induced magnetization: the inverse Faraday effect~\cite{PhysRevLett.15.190,Berritta2016,Mondal_2017,PhysRevB.110.094302,PhysRevB.103.205417}), which can shift the band structure and tune the strength of the altermagnetic order. 
        
        We note that the light-induced magnetization scales quadratically with the amplitude of the light field \cite{PhysRevLett.15.190,Berritta2016}.
        In the \( d_{x^2 - y^2} \)-wave pattern, the RSOC terms are modified by \( \Delta \mathcal{D} \), whereas in the \( d_{xy} \)-wave pattern, a Rashba-Dresselhaus-type SOC emerges. Note that the \( d_{xy} \)-wave pattern corresponds to a 45$^\circ$ rotation of the \( d_{x^2 - y^2} \)-wave pattern in both momentum and spin space.  
		
		For left-handed CPL, we employ $\Omega_{\rm d} \to -\Omega_{\rm d}$, flipping the $x$-component of the vector potential. This reverses the magnetization term proportional to $\Delta$~\cite{PhysRevLett.15.190,PhysRevB.110.094302} and inverts the RSOC anisotropy in $d_{x^2-y^2}$-wave and the Dresselhaus SOC amplitude in $d_{xy}$-wave patterns.
		
		Since RSOC shows stronger anisotropy in the \( d_{x^2 - y^2} \)- than in the \( d_{xy} \)-wave pattern, we focus on \( d_{x^2 - y^2} \)-wave altermagnets for highly tunable spin polarizations. It is convenient to express the effective Hamiltonian as $\mathcal{H}^{{\rm eff},d_{x^2-y^2}}_{\vec{k}}  = d_0(k)\, \sigma_0 + \vec{d}(\vec{k}) \cdot \vec{\sigma}$, with eigenenergies $\varepsilon_{\pm \vec{k}} = d_0(k) \pm |\vec{d}(\vec{k})|$ and eigenstates $|u_{\pm \vec{k}} \rangle$ describing spin-split bands. Figures~\ref{f1}(b)--(d) show the evolution of spin textures, $\langle u_{\pm \vec{k}}| \sigma_z |u_{\pm \vec{k}} \rangle$, in a \( d_{x^2 - y^2} \)-wave altermagnet as time-reversal and inversion symmetries are sequentially broken: pristine, gated/RSOC, and under CPL. 
        
        In Fig.~\ref{f1}(b), without RSOC or optical driving, spin-split Fermi contours~(corresponding to out-of-plane spins) are antisymmetric under $C_4$ rotations. Time-reversal symmetry breaking remains, giving zero net magnetization despite nontrivial momentum-resolved spin textures. In Fig.~\ref{f1}(c), RSOC breaks inversion symmetry, distorting contours while preserving $C_{4}$ symmetry. In Fig.~\ref{f1}(d), CPL also breaks time-reversal symmetry, inducing net $z$-polarized magnetization, enhancing spin contrast, and driving Floquet band reconstruction. 
		
		While CPL in combination with gating generates anisotropic in-plane spin textures~\cite{yarmohammadi2025anisotropiclighttailoredrkkyinteraction}, CPL alone does not create in-plane spin polarizations. Adding in-plane electric fields completes the scheme for engineering spin polarization across multiple components. These polarizations are mainly governed by the Edelstein effect, as noted in the introduction.
		
		\blue{\textit{Edelstein effect.}}---In non-centrosymmetric systems with SOC, an applied electric field can induce a nonequilibrium spin or orbital polarization—known respectively as the spin Edelstein effect and orbital Edelstein effect or Rashba-Edelstein effect~\cite{edelstein1990spin,Salemi2019,Johansson_2024}. Within linear response theory, the expectation value of an observable $\hat{O}^\ell$~($\ell \in \{x,y,z\}$) as a linear response to an in-plane electric field \( E^j\) along direction \( j \in \{x,y\}\) is given by $\langle \hat{O}^\ell \rangle = \sum_j \chi_{\ell j}^{  } E^j$, where $\chi_{\ell j}^{}$ is the Edelstein response tensor~(susceptibility). For the spin Edelstein effect, $\hat{O}^\ell = \hbar \sigma_\ell / 2$ (spin operator), while for the orbital Edelstein effect, $\hat{O}^\ell = \hat{L}^\ell$ (orbital angular momentum, see below). Using the Kubo formula, $\chi_{\ell j}^{ }$ reads as~\cite{PhysRevMaterials.5.074407,PhysRevMaterials.6.095001}\begin{align}\label{eq_4}
			\chi_{\ell j}^{ } = {} &\frac{ie}{m_{\rm e}} \int \frac{d^2k}{(2\pi)^2} \Bigg(
			\sum_{n} \frac{\partial f_{n\vec{k}}}{\partial \varepsilon_{n\vec{k}}} \frac{O^\ell_{nn,\vec{k}} p^j_{nn,\vec{k}} }{i\tau_{\text{intra}}^{-1}} \nonumber \\
			& - \sum_{n \ne m} \frac{f_{n\vec{k}} - f_{m\vec{k}}}{\varepsilon_{n\vec{k}} - \varepsilon_{m\vec{k}}} 
			\frac{O^\ell_{mn,\vec{k}} p^j_{nm,\vec{k}} }{\varepsilon_{m\vec{k}} - \varepsilon_{n\vec{k}} + i\tau_{\text{inter}}^{-1}} \Bigg)\, ,
		\end{align}where $n = \pm$ is the spin band index, $f_{n\vec{k}}$ is the corresponding Fermi-Dirac distribution, $\tau_{\text{intra}}$~($\tau_{\text{inter}}$) is the intraband~(interband) lifetime, $O^\ell_{mn,\vec{k}} = \langle u_{m\vec{k}} | \hat{O}^\ell | u_{n\vec{k}} \rangle$, and $p^j_{nm,\vec{k}} = m_{\rm e} \langle u_{n\vec{k}} | \hat{v}^j | u_{m\vec{k}} \rangle$, where $\hat{v}^j = \hbar^{-1} \partial \mathcal{H}^{{\rm eff},d_{x^2-y^2}}_{\vec{k}}/\partial_{k_j}$ is the velocity operator. In Eq.~\eqref{eq_4}, the first term reflects intraband contributions, and the second captures interband coherence. 
		
		There are two ways to compute induced responses. One way starts from the unperturbed Hamiltonian and treats both light and in-plane fields perturbatively, requiring linear and quadratic corrections to the electric field—often cumbersome. Our simpler approach uses the dressed Hamiltonian, already incorporating CPL effects, on which we directly compute the Edelstein responses.

        Our model applies only at temperatures well below the energy cutoff of the effective theory; at higher $T$, thermal excitations may access neglected high-energy scales. We thus focus on low $T$, where the approximation is valid. As $T \to 0$, $\partial f / \partial \varepsilon \to -\delta(\varepsilon - \mu)$, with $\mu$ the chemical potential (tunable within spin bands, analogous to doping), yielding the intraband contribution\begin{equation}
        	\chi^{\text{intra}}_{\ell j} = -\frac{e \tau_{\text{intra}}}{m} \int \frac{d^2k}{(2\pi)^2} \sum_n \delta(\varepsilon_{n\vec{k}} - \mu)\, O^\ell_{nn,\vec{k}}\, p^j_{nn,\vec{k}}\, .
        \end{equation}
		
		While the atom-centered approximation is commonly employed to study the orbital Edelstein effect in bulk magnetic materials and full-band models~\cite{Johansson_2024,Atencia31122024,PhysRevB.94.121114}, in few-band models—particularly two-band—the orbital moment can be expressed as~\cite{PhysRevB.53.7010,PhysRevB.102.235426}
		\begin{equation}
			\vec{L}_{mn\vec{k}} = i \langle \nabla_{\vec{k}} u_{m\vec{k}} | \times \big[\mathcal{H}^{{\rm eff},d_{x^2-y^2}}_{\vec{k}} - \varepsilon_{n\vec{k}}\big] | \nabla_{\vec{k}} u_{n\vec{k}} \rangle\, .
		\end{equation}This formulation captures the internal angular momentum of Bloch wavepackets, which is primarily governed by the Berry curvature of the bands.
		
		The complete spin and orbital ingredients of the Edelstein effect are presented in Appendix~\hyperlink{mylinkB}{B}.
		
		{\blue{\textit{Results and discussion.}}---In the following, we first consider an ideal $d_{x^2 - y^2}$ altermagnet with $\mathcal{D}=1$ and weak gate-induced RSOC $\lambda=0.2$ eV$\cdot$\AA. We then tune $\mathcal{D}$ and $\lambda$ to study responses to light potentials $\Delta$ and in-plane fields $(E^x,E^y)$. Two doping levels are set by $\mu = \varepsilon_{\pm} = \varepsilon_{\pm,\vec{k}=0}$, the band edges at $\vec{k}=0$, where state degeneracy and spin polarizations peak in the pristine phase. We note that when $\mu$ lies within spin bands, the main conclusions remain valid. We use $k_{\rm B}T \approx 1$ meV and normalize the Edelstein susceptibility by $\chi_0=e/4\pi^2$ with natural units $e=\hbar=m_{\rm e}=1$. Intraband and interband lifetimes $\tau_{\rm intra}^{-1}$ and $\tau_{\rm inter}^{-1}$ are set to $0.5$ eV, an acceptable value for metallic systems~\cite{PhysRevMaterials.6.095001,Salemi2019}, noting longer lifetimes enhance spin polarizations but do not alter qualitative trends. We also emphasize that numerical and analytical results (Appendix~\hyperlink{mylinkB}{B}) agree perfectly, thus, duplicate plots are omitted. We use $\Delta>0$ for right-handed and $\Delta<0$ for left-handed CPL.
		
		Owing to the $d_{x^2-y^2}$-wave symmetry of our model, the generation of orbital polarizations—both in-plane and out-of-plane—is negligible with and without gating or CPL, as the orbital responses to in-plane electric fields cancel out with opposite signs along different momentum directions, as confirmed by the orbital results shown in Fig.~\ref{f2}.
		
		Under CPL illumination, spin polarization arises along $z$. We focus on the Edelstein effect to study in-plane polarizations induced by $x$- and $y$-directed electric fields. By mirror symmetry in the $z$ direction (reflection about the plane), the in-plane Edelstein effect cannot generate $z$-polarization (Fig.~\ref{f2}), but it induces in-plane polarizations that are strongly reshaped anisotropically under combined gating and optical driving.
			
		While we focus on the dominant off-diagonal (transverse) susceptibility components, the longitudinal components ($\chi_{xx}$ and $\chi_{yy}$) show similar behavior (dark gray and yellow curves in Fig.~\ref{f2}(a) and~\ref{f2}(b)).\begin{figure}[t]
			\centering
			\includegraphics[width=0.9\linewidth]{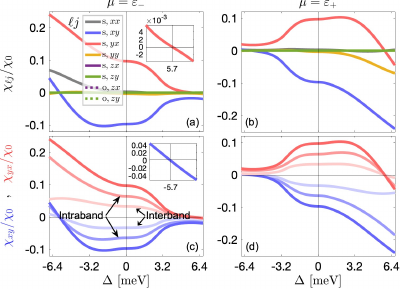}
			\caption{Calculated spin (‘s’) and orbital (‘o’) Edelstein susceptibilities $\chi_{ij}/\chi_0$ ($\chi_0 = e/4\pi^2$) versus light potential $\Delta$ under CPL for an ideal altermagnet with $\mathcal{D}=1$ and weak RSOC $\lambda=0.2$ eV$\cdot$\AA. Panels~(a,b): chemical potentials $\mu = \varepsilon_\mp$ (band edges at $\vec{k}=0$). Panels~(c,d) separate intra- and interband contributions. Without light ($\Delta=0$), in-plane fields yield $\chi_{xy}$ and $\chi_{yx}$ of opposite sign. With CPL ($\Delta \neq 0$), anisotropic susceptibilities, large tunable spin polarizations, and chiral optical activity emerge, demonstrating directional control. At $\Delta_{\rm c}\!\approx \!\pm 5.7$ meV (insets), in-plane spin polarizations vanish and subsequently reverse as the spin susceptibilities change sign. This highlights the synergy of Edelstein and Floquet mechanisms in achieving highly tunable spin responses in Rashba altermagnets. Note that all $\chi_{z j}$ components in (a,b) coincide at zero, causing the dashed lines to lie directly on top of the solid lines.}
			\label{f2}
		\end{figure}
			
			Figure~\ref{f2} reveals the light-induced evolution of the spin and orbital Edelstein effects as functions of the light potential $\Delta$ in a 2D $d_{x^2-y^2}$-wave Rashba altermagnet. In the absence of light (\(\Delta= 0\)), the inherent spin textures enforce antisymmetric susceptibilities independent of doping level:{\small\begin{align}\label{eq_6}
					\chi_{xy}^{ }(\Delta = 0, \mu \neq 0) = - \chi_{yx}^{ }(\Delta = 0, \mu \neq 0) \mapsto \langle \hat{S}^x\rangle = - \langle \hat{S}^y\rangle\, .
			\end{align}}This antisymmetry reflects the time-reversal and inversion symmetries inherent to the undriven system. Finite $\langle \hat{S}^x \rangle$ and $\langle \hat{S}^y \rangle$ arise under in-plane electric fields, producing antisymmetric $x$- and $y$-polarizations. Combined with CPL, which induces out-of-plane $z$-polarization, this yields full directional control of spin polarization in altermagnets. Without CPL, regardless of doping energy in spin-up or spin-down bands, we find\begin{align}\begin{pmatrix}
					\langle \hat{S}^x\rangle\\
					\langle \hat{S}^y\rangle
				\end{pmatrix} = \begin{pmatrix}
					\chi_{xx} &\chi_{xy}\\
					-\chi_{xy} &-\chi_{xx}
				\end{pmatrix}\begin{pmatrix}
					E^x\\
					E^y
				\end{pmatrix}\,.
			\end{align} 
			
			We demonstrate the combined effect of CPL and in-plane electric fields with weak gating $\lambda = 0.2$ eV$\cdot$\AA. The antisymmetry in Eq.~\eqref{eq_6} is broken by the light helicity ($\Delta \neq 0$), producing anisotropic spin susceptibilities. Depending on helicity, $\langle \hat{S}^x \rangle$ and $\langle \hat{S}^y \rangle$ can be selectively tuned or switched. Without gating, this tunability is unattainable (Eq.~\eqref{eq_2a}), reflecting reciprocity breakdown, $\langle \hat{S}^x\rangle \neq - \langle \hat{S}^y\rangle$. Figures~\ref{f2}(c)--(d) separate total $\chi_{xy}$ and $\chi_{yx}$ into intra- and interband contributions; the intraband part dominates, showing that polarization arises from light-modified band velocities and spin textures rather than virtual transitions.
			
			In particular, the induced spin polarizations along $x$ and $y$, given by $\chi_{xy}$ and $\chi_{yx}$, vanish at critical light potentials (insets of Fig.~\ref{f2}), depending on CPL helicity. From modified RSOC components in Eq.~\eqref{eq_2a}, $(\lambda_x,\lambda_y) = (\lambda - \frac{\hbar^2 \Delta \mathcal{D}}{m \lambda}, \lambda + \frac{\hbar^2 \Delta \mathcal{D}}{m \lambda})$, $\chi_{xy}$ vanishes under RCPL at $\lambda_x=0$, while $\chi_{yx}$ vanishes under LCPL at $\lambda_y=0$. The critical light potential is then\begin{align}\label{eq_7}
				\Delta^\pm_{\rm c} \approx \pm \frac{m\,\lambda^2}{\hbar^2\,\mathcal{D}}\, ,
			\end{align}where $+$ and $-$ stand for right and left light helicity, respectively. For $\hbar \Omega_{\rm d} \approx 1$ eV and $\mathcal{D}=1$, the critical light-induced exchange potential is $\Delta_{\rm c} \approx \pm 5.7$ meV (Fig.~\ref{f2}). At this value, in-plane spin polarizations switch off and then reverse due to a sign change in spin susceptibilities. The corresponding critical vector potential is $\mathcal{A}_{\rm c} \approx \pm 0.377$ \AA$^{-1}$ and critical field $E_0^{\rm c} \approx 3.7$ V/nm, independent of $\mu$. While experimentally challenging, these values are achievable with current ultrafast lasers~\cite{Higuchi2017}, though real setups involve losses not captured in a zero-temperature free-standing model~\cite{Higuchi2017,liu2024evidencefloquetelectronicsteady}.
			
			Following the analysis of the spin Edelstein components at finite doping ($\mu \neq 0$) in Fig.~\ref{f2}, when light is present ($\Delta\neq 0$), we find\begin{subequations}\label{eq_8}
				\begin{align}
					\chi_{xy}^{}(\Delta < 0, \mu = \varepsilon_-) = {} &- \chi_{yx}^{}(\Delta > 0, \mu = \varepsilon_+)\, ,\\
					\chi_{xy}^{} (\Delta > 0, \mu = \varepsilon_-) = {} &- \chi_{yx}^{}(\Delta < 0, \mu = \varepsilon_+)\, ,
				\end{align}
			\end{subequations}revealing a chiral antisymmetry between the spin-up and spin-down bands. This antisymmetry stems from inversion symmetry of spin components and light helicity acting oppositely on different spins, a phenomenon we term the \textit{chiral optical activity}. This behavior is unique to altermagnets, as non-altermagnetic systems lack the parameter $\mathcal{D}$ in light-modified RSOCs.
			
			The full control mechanism, combining gating, CPL, and in-plane electric fields, becomes now evident. In a nutshell, the effects of RCPL and LCPL~($\sigma^+$ for RCPL and $\sigma^-$ for LCPL) when doping spin band~($\sigma$ for spin-up and $-\sigma$ for spin-down) are given by the following chiral features\begin{align}
				\begin{pmatrix}
					\langle \hat{S}^x\rangle\\
					\langle \hat{S}^y\rangle
				\end{pmatrix} = {} & \begin{pmatrix}
					\chi_{xx}^{}(\sigma^{\pm},\sigma) &\chi_{xy}^{}(\sigma^{\pm},\sigma)\\
					-\chi_{xy}^{}(\sigma^{\mp},-\sigma) &-\chi_{xx}^{}(\sigma^{\mp},-\sigma)
				\end{pmatrix}\begin{pmatrix}
					E^x\\
					E^y
				\end{pmatrix}\,,
		\end{align}
     enabling highly tunable in-plane spin polarizations in altermagnets. We stress that the $d_{x^2-y^2}$-wave symmetry in altermagnets forbids out-of-plane ($z$) spin polarization via the Edelstein effect, which can only be induced directly by CPL. \begin{figure}[t]
				\centering
				\includegraphics[width=0.85\linewidth]{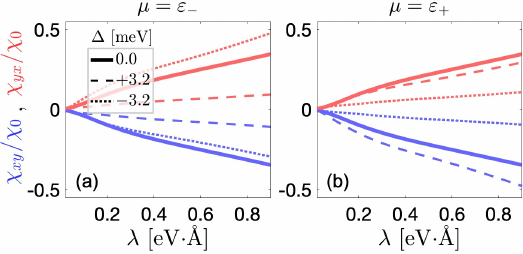}
				\caption{Calculated transverse spin susceptibilities $\chi_{xy}/\chi_0$ (blue) and $\chi_{yx}/\chi_0$ (red) versus RSOC $\lambda$ in an ideal altermagnet ($\mathcal{D}=1$) for various light potentials: $\Delta = 0$ (solid), $+3.2$ meV (dashed), and $-3.2$ meV (dotted). Panel (a): $\mu = \varepsilon_-$ at $\vec{k} = 0$ for the spin-down band; panel (b): $\mu = \varepsilon_+$ at $\vec{k} = 0$ for the spin-up band. Chiral optical activity from combined light and in-plane fields is independent of finite RSOC.}
				\label{f3}
				\end{figure}
			
			Next, we examine how gate voltage, or gate-induced RSOC, affects spin Edelstein polarizations (Fig.~\ref{f3}). Susceptibilities increase with $\lambda$, enhancing polarizations. Without light ($\Delta = 0$), the system retains antisymmetric spin susceptibility, $\chi_{xy} = -\chi_{yx}$, confirmed by solid curves in Fig.~\ref{f3}. Introducing CPL ($\Delta \neq 0$) breaks this antisymmetry for all $\lambda$, depending on light helicity and doping.
			
			At $\mu = \varepsilon_-$ (Fig.~\ref{f3}(a)), RCPL ($\Delta = +3.2$~meV, dashed) largely preserves antisymmetry, like the undriven case. In contrast, LCPL ($\Delta = -3.2$~meV, dotted) induces pronounced anisotropy between $\chi_{xy}$ and $\chi_{yx}$, showing helical light-spin coupling enhancing one spin texture. At $\mu = \varepsilon_+$ (Fig.~\ref{f3}(b)), the effect is similar but with reversed curvature, reflecting opposite upper-band spin polarization. The susceptibility components no longer mirror each other under CPL, evidencing breakdown of response reciprocity with helicity reversal (cf. Fig.~\ref{f2}).\begin{figure}[b]
				\centering
				\includegraphics[width=0.85\linewidth]{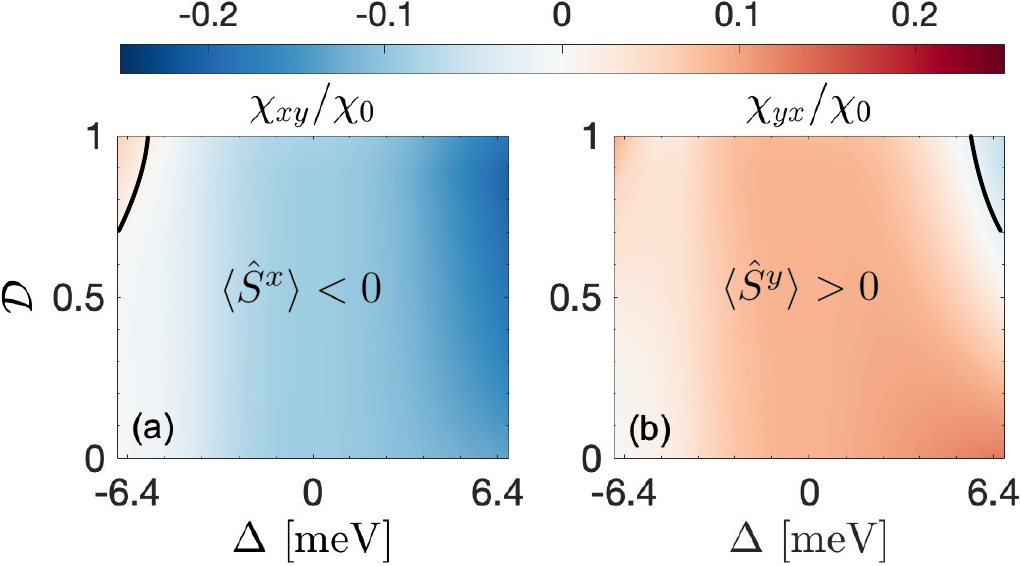}
				\caption{Calculated spin susceptibilities $\chi_{xy}$ (a) and $\chi_{yx}$ (b) for doping into the spin-up band ($\mu = \varepsilon_+$) with weak RSOC $\lambda = 0.2$ eV$\cdot$\AA. For spin-down doping ($\mu = \varepsilon_-$), the behaviors reverse. Both components are plotted versus light potential $\Delta$ and altermagnetism $\mathcal{D}$, illustrating highly tunable CPL-induced spin polarizations and the critical phase boundary where polarizations switch.}
				\label{f4}
			\end{figure}
			
			Notably, Eq.~\eqref{eq_6} implies that the crossing point at which \(\chi_{xy}^{}\) and \(\chi_{yx}^{}\) vanish and spin polarization is reversed, remains fixed and is not shifted by RSOC. These observations across varying RSOC strengths reinforce our claim regarding the robustness of the chiral optical responses.
			
			To highlight the sensitivity of spin polarization chirality to altermagnetic strength, we plot $\chi_{xy}/\chi_0$ and $\chi_{yx}/\chi_0$ versus $\Delta$ and $\mathcal{D}$ (Figs.~\ref{f4}(a,b)). CPL ($\Delta \neq 0$) breaks intrinsic reciprocity, seen in anisotropic color distributions, signaling spin antisymmetry breakdown under helicity reversal. For small $\mathcal{D}$, $\chi_{xy}$ and $\chi_{yx}$ remain nearly antisymmetric about $\Delta=0$, showing that strong spin anisotropy is essential for nonreciprocal features. Solid black lines mark sign-change boundaries, indicating possible topological spin texture transitions induced via RSOC. Depending on $\mathcal{D}$, spin polarization orientation can be switched by tuning $\Delta$; e.g., at $\mathcal{D}=1$ in the spin-up band, $x$~($y$)-polarization reverses at $\Delta \approx -5.7$~meV~($\Delta \approx +5.7$~meV), with corresponding reversal in the spin-down band.
			
			\blue{\textit{Experimental perspective.}}---The out-of-plane spin polarization $\langle \hat{S}^z \rangle$ is set by the light potential $\Delta$, achievable with current laser technology. In natural units ($e = \hbar = m_{\rm e} = 1$), $\mu_{\rm B} = 1/2$, and using $\rho(\mu) \approx 1$--$10$ states/eV/unit cell per spin, $\langle S_z \rangle = (1/2)\, \rho(\mu)\, \Delta$ gives, for $a \approx 4$~\AA~(a representative lattice constant) and $\Delta = 6.5$ meV,\begin{align}
					\langle \hat{S}^z \rangle \approx 10.4 \times 10^{-2}\, \mu_{\rm B}.
				\end{align}For in-plane polarizations $\langle \hat{S}^{x,y} \rangle$, the velocity term in Eq.~\eqref{eq_4} yields a momentum derivative unit of 0.4 nm. Restoring $\mu_{\rm B}$ and lattice units, for weak gating ($\lambda = 0.2$ eV$\cdot$\AA), in-plane fields $(E^x,E^y)=(1,1)$ V/nm, and $|\Delta| \approx 6.5$ meV, we find\begin{align}
				\langle \hat{S}^{x} \rangle \approx 4.32 \times 10^{-2} \,\mu_{\rm B},\quad\langle \hat{S}^{y} \rangle \approx 9.6 \times 10^{-2} \,\mu_{\rm B}\, .
			\end{align}These magnitudes are experimentally detectable~\cite{PhysRevLett.119.087203} and can be increased by stronger gating (Fig.~\ref{f3}) or purer materials with smaller inverse lifetimes than 0.5 eV.
		
			For experimental setups, one can use heterostructures of thin films grown epitaxially on substrates (e.g., TiO$_2$ for RuO$_2$~\cite{doi:10.1126/sciadv.adj4883}) to achieve 2D-like behavior, with top/bottom gates (e.g., high-$\kappa$ dielectrics) inducing RSOC in van der Waals heterostructures. Circularly polarized mid-infrared or terahertz pulses ($\Omega_\text{d} \approx 1$ eV, $E_0 \approx 3$--$5$ V/nm~\cite{Higuchi2017}) enable Floquet engineering~\cite{plouff2024revisitingaltermagnetismruo2study}. Spin textures and symmetry breaking can be probed via TR-ARPES, while the Edelstein effect is measurable through Hall-bar transport or spin pumping. Chiral responses are accessible via MOKE or Faraday rotation near band edges~\cite{RevModPhys.82.2731,PhysRevB.106.014410}. Despite challenges such as material quality and ultrafast timescales, recent advances demonstrate that these measurements are feasible.
			
			\blue{\textit{Summary.}}---We have shown that combining gating, optical driving, and in-plane electric fields enables versatile control of spin polarization in $d$-wave altermagnets. Starting from the dressed Hamiltonian that includes CPL and gating, induced in-plane responses can be computed directly, avoiding complex perturbative corrections. Optical driving generates out-of-plane polarization, while in-plane fields induce in-plane polarizations via the Edelstein effect, all within experimentally accessible ranges. This multi-field approach allows highly tunable and switchable spin polarizations in all directions. Additionally, we find that selective doping of spin bands produces chiral optical activity, which highlights a feature unique to altermagnets. Our results provide a practical route for complete spin control in altermagnets, paving the way for advanced spintronic applications.
			
			\blue{\textit{Acknowledgments.}}---M.Y.\ gratefully acknowledges the hospitality of the Max Planck Institute for the Physics of Complex Systems and of Uppsala University during his visit, where parts of this work were carried out. M.Y.\ was supported by the Department of Energy, Office of Basic Energy Sciences, Division of Materials Sciences and Engineering under Contract No.\ DE-FG02-08ER46542 for the formal developments, the analytical/numerical work, and the writing of the manuscript. J.L.\ was supported by the Research Council of Norway through its Centres of Excellence funding scheme Grant No.\ 262633 and Grant No.\ 353894. M. Berritta\ and P.M.O.\ acknowledge support by the Swedish Research Council (VR), the German Research Foundation (Deutsche Forschungsgemeinschaft) through CRC/TRR 227 “Ultrafast Spin Dynamics” (project MF, project-ID: 328545488), and the K.\ and A.\ Wallenberg Foundation (Grants No.\ 2022.0079 and 2023.0336). Part of the calculations were supported by the National Academic Infrastructure for Supercomputing in Sweden (NAISS) at NSC Link\"oping, funded by VR through Grant No.\ 2022-06725. M. Bukov\ was funded by the European Union (ERC, QuSimCtrl, 101113633). L. Š.\ acknowledges funding from the ERC Starting Grant No. 101165122.

		\twocolumngrid
		\bibliography{bib.bib}
		{\allowdisplaybreaks
			\onecolumngrid
			\subsection{\large End Matter}\label{ap1}
			\hypertarget{mylinkA}{\blue{\textit{Appendix A}}: \blue{\textit{Time-dependent Hamiltonian.}}}---In this Appendix, we provide the time-dependent Rashba Hamiltonians in the minimal coupling regime:	{\small\begin{subequations}
					\begin{align}
						\mathcal{H}^{{\rm R},d_{x^2-y^2}}_{\vec{k} - \tfrac{e}{\hbar}\,\vec{A}(t)}(t) = {} &\tfrac{\hbar^2 (k_x^2 + k_y^2)}{2m_{\rm e}} \sigma_0 
						+\tfrac{\hbar^2 \mathcal{D} (k_x^2 - k_y^2)}{2m_{\rm e}} \sigma_z  + \lambda\, (k_x\,\sigma_y - k_y\,\sigma_x) + \frac{\hbar^2}{2\,m_{\rm e}}\left[\frac{e^2\,A_{\rm d}^2}{\hbar^2} - 2 \frac{e\,A_{\rm d}}{\hbar}\left(k_x\sin(\Omega_{\rm d}t)+k_y\cos(\Omega_{\rm d}t)\right)\right]\sigma_0 \notag \\  {} + & \frac{\hbar^2 \beta}{2\,m_{\rm e}}\left[-\frac{e^2\,A_{\rm d}^2}{\hbar^2}\cos(2\Omega_{\rm d}t) - 2 \frac{e\,A_{\rm d}}{\hbar}\left(k_x\sin(\Omega_{\rm d}t)-k_y\cos(\Omega_{\rm d}t)\right)\right]\sigma_z - \lambda \frac{e\,A_{\rm d}}{\hbar} \left(\sin(\Omega_{\rm d}t) \sigma_y-\cos(\Omega_{\rm d}t)\sigma_x\right)\,,\\
						\mathcal{H}^{{\rm R},d_{xy}}_{\vec{k} - \tfrac{e}{\hbar}\,\vec{A}(t)}(t) = {} &\tfrac{\hbar^2 (k_x^2 + k_y^2)}{2m_{\rm e}}  \sigma_0 + \tfrac{\hbar^2 \widetilde{\mathcal{D}} \,k_x\, k_y}{m_{\rm e}}\sigma_z  + \lambda\, (k_x\,\sigma_y - k_y\,\sigma_x) +\tfrac{\hbar^2 \beta}{m_{\rm e}}\left[\tfrac{e^2\,A_{\rm d}^2}{2 \hbar^2}\sin(2\,\Omega_{\rm d}\,t)- \tfrac{e\,A_{\rm d}}{\hbar}\left(k_x\cos(\Omega_{\rm d}t)+k_y\sin(\Omega_{\rm d}t)\right)\right]\sigma_z \notag \\ {} - & \lambda \frac{e\,A_{\rm d}}{\hbar} \left(\sin(\Omega_{\rm d}t) \sigma_y-\cos(\Omega_{\rm d}t)\sigma_x\right)\, (k_x\,\sigma_y - k_y\,\sigma_x)
			\end{align}\end{subequations}}Using $\mathcal{H}^{\rm F}_0 = \frac{1}{T} \int^T_0 \mathcal{H}^{\rm R}_{\vec{k} - \tfrac{e}{\hbar}\,\vec{A}(t)}(t) dt$ and $\mathcal{H}^{\rm F}_{\pm 1} = \frac{1}{T} \int^T_0 \mathcal{H}^{\rm R}_{\vec{k} - \tfrac{e}{\hbar}\,\vec{A}(t)}(t) e^{\pm i \Omega_{\rm d} t} dt$, the zeroth-order $\mathcal{H}^{\rm F}_0$ and first-order $\mathcal{H}^{\rm F}_{\pm1}$ Floquet terms are obtained. Finally, in the van Vleck expansion $\mathcal{H}^{\rm eff}_{\vec{k}} \approx \mathcal{H}_0^{\rm F} + \frac{[\mathcal{H}^{\rm F}_{-1}, \mathcal{H}^{\rm F}_{+1}]}{\hbar\Omega_{\rm d}}$, the effetive time-independent Hamiltonian is calculated.
			
			\hypertarget{mylinkB}{\blue{\textit{Appendix B}}: \blue{\textit{Spin, orbital, and momentum operators.}}}---In this Appendix, we outline the complete set of ingredients required to evaluate the spin and orbital Edelstein response functions, formulated as follows:\begin{subequations}
				\begin{align}
					S^{x}_{\pm \pm,\vec k} = {} &\pm \frac{\hbar}{2} \sin(\theta_{\vec k}) \cos(\phi_{\vec k})\, ,\quad
					S^{y}_{\pm \pm,\vec k} = {} \pm \frac{\hbar}{2} \sin(\theta_{\vec k}) \sin(\phi_{\vec k})\, ,\quad
					S^{z}_{\pm \pm,\vec k} = {} \pm \frac{\hbar}{2} \cos(\theta_{\vec k}) \, ,\\
					p^{x}_{\pm \pm,\vec k} = {} & \hbar k_x \mp \frac{\lambda_x\,m_{\rm e}}{\hbar} \sin(\theta_{\vec k}) \cos(\phi_{\vec k}) \pm \hbar\mathcal{D} k_x \cos(\theta_{\vec k})\, ,\label{eq_B1d}\\
					p^{y}_{\pm \pm,\vec k} = {} &\hbar k_y \pm \frac{\lambda_y\,m_{\rm e}}{\hbar} \sin(\theta_{\vec k}) \sin(\phi_{\vec k}) \mp \hbar\mathcal{D}  k_y \cos(\theta_{\vec k})\, ,\label{eq_B1e}\quad
					p^{z}_{\pm \pm,\vec k} = {} 0\, ,\\
					S^{x}_{\pm \mp,\vec k} = {} &\frac{\hbar}{2}\left(\cos^2(\theta_{\vec k}/2)e^{\pm i\phi_{\vec k}}-\sin^2(\theta_{\vec k}/2)e^{\mp i\phi_{\vec k}}\right)\, ,\\
					S^{y}_{\pm \mp,\vec k} = {} &\frac{\mp i\hbar}{2}\left(\sin^2(\theta_{\vec k}/2)e^{\mp i\phi_{\vec k}}+\cos^2(\theta_{\vec k}/2)e^{\pm i\phi_{\vec k}}\right)\, ,\quad
					S^{z}_{\pm \mp,\vec k} = {} -\frac{\hbar}{2}\sin(\theta_{\vec k})\, ,\\
					p^{x}_{\pm \mp,\vec k} = {} &-\frac{\lambda_x m_{\rm e}}{\hbar}\left(\cos^2(\theta_{\vec k}/2)e^{\pm i\phi_{\vec k}}-\sin^2(\theta_{\vec k}/2)e^{\mp i\phi_{\vec k}}\right)- \hbar \mathcal{D}  k_x \sin(\theta_{\vec k})\, ,\\
					p^{y}_{\pm \mp,\vec k} = {} &\mp i\frac{\lambda_y m_{\rm e}}{\hbar}\left(\cos^2(\theta_{\vec k}/2)e^{\pm i\phi_{\vec k}}+\sin^2(\theta_{\vec k}/2)e^{\mp i\phi_{\vec k}}\right)+ \hbar \mathcal{D}  k_y \sin(\theta_{\vec k})\, ,\quad
					p^{z}_{\pm \mp,\vec k} = {} 0\, ,
				\end{align}
			\end{subequations}where\begin{align}
				\cos\theta_{\vec k}=\frac{d_z(\vec{k})}{d(\vec{k})},\quad
				\tan\phi_{\vec k}=\frac{d_y(\vec{k})}{d_x(\vec{k})}\,,
			\end{align}stemming from the Hamiltonian $\mathcal{H}^{{\rm eff},d_{x^2-y^2}}_{\vec{k}}  = d_0(k)\, \sigma_0 + \vec{d}(\vec{k}) \cdot \vec{\sigma}$ in Eq.~\eqref{eq_2a} with eigenenergies $\varepsilon_{\pm \vec{k}} = d_0(k) \pm \, |\vec{d}(\vec{k})|$ and eigenstates\begin{align}
				|u_{+\vec{k}}\rangle=\begin{pmatrix}\cos\frac{\theta_{\vec k}}{2}\,e^{-i\phi_{\vec k}}\\\\\sin\frac{\theta_{\vec k}}{2}\end{pmatrix},\quad
				|u_{-\vec{k}}\rangle=\begin{pmatrix}-\sin\frac{\theta_{\vec k}}{2}\,e^{-i\phi_{\vec k}}\\\\\cos\frac{\theta_{\vec k}}{2}\end{pmatrix}.
			\end{align}Moreover, in two dimensions, the orbital magnetic moment \( m^\ell_{mn\vec{k}} = \mu_{\rm B}	L^\ell_{mn\vec{k}}\), with \( \mu_{\rm B} = e\hbar / (2m_{\rm e}) \) as the Bohr magneton and $\ell \in \{x,y,z\}$, connects to the Berry curvature \( \Omega^\ell_{mn}(\vec{k}) \). Notably, within the model under consideration, interband transitions do not contribute to the orbital Edelstein effect due to the vanishing of off-diagonal matrix elements: \( \langle u_{\pm \vec{k}}|u_{\mp \vec{k}}\rangle = 0 \). Thus, for $m = n$ intraband contributions, we write
			\begin{align}
				L^\ell_{nn\vec{k}} = -\frac{e}{\hbar \mu_{\rm B}} \, \varepsilon_{n\vec{k}} \, \Omega^\ell_{nn}(\vec{k})\,,\quad
				\Omega^\ell_{nn}(\vec{k}) = \frac{n\,\epsilon_{\alpha \beta \ell}}{2} \frac{\vec{d}(\vec{k})\cdot\left(\partial_{k_\alpha}\vec{d}(\vec{k}) \times\partial_{k_\beta}\vec{d}(\vec{k})\right)}{|\vec{d}(\vec{k})|^3}\, ,
			\end{align}where \( \epsilon_{\alpha\beta\ell} \) is the Levi-Civita tensor. As there is no $k_z$ term in the Hamiltonian, we have $L^x_{nn\vec{k}} = L^y_{nn\vec{k}} = 0$, yielding an angular momentum (in dimensionless units) only along the $z$-direction:\begin{align}\label{eq_13}
					L^z_{n n\vec{k}} = \frac{n\,e\,\varepsilon_{n\vec{k}}  \lambda_x\,\lambda_y\left(\tfrac{\hbar^2 \mathcal{D}(k^2_x - k^2_y)}{2\,m_{\rm e}}-\Delta \right)}{2 \hbar \mu_{\rm B} \left(d_x^2(\vec{k}) + d_y^2(\vec{k}) + d_z^2(\vec{k})\right)^{3/2}}\,.
			\end{align}
		}
	\end{document}